\documentclass[aps,twocolumn,prb]{revtex4}
\usepackage{multirow}
\usepackage{graphicx}
\usepackage{amsmath}
\usepackage{enumerate}
\usepackage{epstopdf}
\begin{document}
\title{Structure factor of low-energy spin excitations in a $S=1/2$ kagome antiferromagnet}
\author{Zhihao Hao}
\author{Oleg Tchernyshyov}
\affiliation{Department of Physics and Astronomy, Johns Hopkins
University, Baltimore, Maryland 21218, USA}

\begin{abstract}
The ground state of the $S=1/2$ Heisenberg antiferomagnet on kagome can be viewed as a collection of fermionic spinons bound into small, heavy singlet pairs. Low-energy magnetic excitations in this system correspond to breaking the pairs into individual spinons. We calculate the structure factor for inelastic neutron scattering from independent spinon pairs.
\end{abstract}

\maketitle

\section{Introduction}

\begin{figure}
\includegraphics[width=0.9\columnwidth]{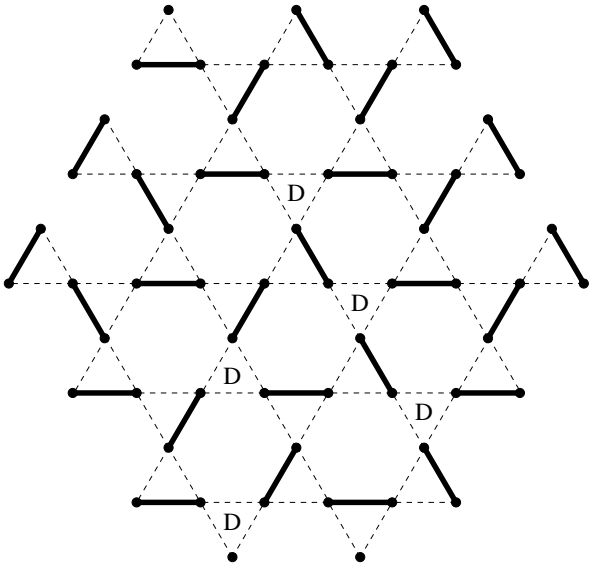}
\caption{A dimer covering of kagome.  Defect triangles (D) lack dimers.}
\label{fig-lattice}
\end{figure}

The spin-1/2 Heisenberg antiferromagnet on kagome (Fig.~\ref{fig-lattice}) has attracted attention of theorists for more than two decades\cite{PhysRevLett.62.2405, PhysRevB.48.13647}.  Synthesis of new magnetic materials with Cu$^{2+}$ spins forming precisely such a lattice has generated renewed interest in this system.\cite{science.321.1306, nature.464.199} Strong frustration, induced by the non-bipartite geometry, and strong quantum fluctuations suppress long-range magnetic order and open intriguing possibilities of an exotic quantum-disordered ground state and unusual magnetic excitations. Numerical calculations\cite{EuroPhysJB.2.501,jiang:117203} suggest that the ground state of the system is a total spin singlet and that magnetic excitations are gapped. Among the proposed ground states of this system are a valence-bond crystal (VBC)\cite{singh:180407,nikolic:024401,PhysRevB.63.014413} breaking lattice symmetries or a spin liquid of some sort.\cite{ryu:184406,ran:117205,hermele:224413}  Recent synthesis and experimental characterization of herbertsmithite ZnCu$_3$(OH)$_3$Cl$_2$,\cite{helton:107204,mendels:077204,ofer-2006,imai:077203,shlee-2007,vries:157205,jpcs.145.012002,jpcs.145.012004,olariu:087202,PhysRevLett.103.237201,Helton:10021091} where Cu$^{2+}$ ions carry $S=1/2$ and form a perfect kagome
lattice, provides additional motivation for theoretical studies of this model.

The simplest model takes into account exchange interactions between nearest neighbors represented by the Hamiltonian
\begin{equation}
H = J\sum_{<ij>}\mathbf{S}_{i}\cdot\mathbf{S}_{j},
\label{eq:H}
\end{equation}
Recasting the Hamiltonian as $J\sum_\triangle S^2_\triangle/2$, where $\mathbf S_\triangle$ is the net spin of triangle $\triangle$, suggests a plausible route to constructing the ground state of the system: minimize the total spin of every triangle, $S_\triangle = 1/2$, by locking two of its three spins in a $S=0$ bond.  Unfortunately, this program fails, as a simple counting argument shows.\cite{PhysRevLett.62.2405}  A kagome with $N$ triangles contains $3N/2$ spins and thus no more than $3N/4$ singlet bonds can be formed, leaving at least one in four triangles without a singlet bond (Fig.~\ref{fig-lattice}).  As a result, a short-range valence-bond state is not a ground state of the Hamiltonian (\ref{eq:H}).

Although the na\"{\i}ve approach to constructing a valence-bond ground state fails, it nonetheless provides a useful starting point by forcing us to regard most of the lattice (triangles with a singlet bond) as the vacuum containing relatively dilute dynamical objects (defect triangles).  Elucidating the properties of an isolated defect triangle is the next logical step towards understanding the physics of the model.  Work in this direction was carried out early on by Elser and Zeng,\cite{PhysRevB.48.13647,PhysRevB.51.8318} who considered an isolated defect triangle on the Husimi cactus\cite{JStatPhys1.27.1237,JPhysA.5.1541}, a Bethe lattice of corner-sharing triangles.  They found that quantum fluctuations are strongly localized in the vicinity of a defect.  By treating the kagome antiferromagnet as a dilute ensemble of fluctuating defects, Elser and Zeng obtained an estimate of its ground-state energy in excellent agreement with numerical diagonalization.

More recently, we pointed out\cite{PhysRevLett.103.187203} that defect triangles are composite objects: they are bound states of two quasiparticles with $S=1/2$.  These quasiparticles bear strong resemblance to spinons of the $\Delta$, or sawtooth, chain\cite{PhysRevB.53.6393,PhysRevB.53.6401}.  The spinons come in two flavors: kinks and antikinks.  Kinks are localized and have zero excitation energy, while antikinks are mobile with the lowest energy of $0.218 J$.  A defect triangle on the Husimi cactus is a tightly bound state of two antikinks with total $S=0$.  Lowest-energy spin excitations correspond to breaking up the pair into two mobile antikinks.  The spin gap is set by the binding energy of the pair, $0.06 J$.  This value, obtained for a single defect on a cactus, agrees with the spin-gap estimates for kagome based on a series expansion\cite{singh:180407} ($0.08 J \pm 0.02 J$) and DMRG\cite{jiang:117203} ($0.055J \pm 0.005 J$).  This agreement suggests that viewing the kagome antiferromagnet as a dilute ensemble of spinon pairs is a useful point of departure for further explorations of this model.

We would like to stress that antikink spinons are not elementary excitations of the kagome antiferromagnet but rather its building blocks.  (In the same way, quarks are building blocks of a baryon but not its elementary excitations.)  One in four triangles carrying a spinon pair translates into 1/3 of an antikink per site.  We have found that antikinks exhibit fermionic statistics.  The minus sign upon an exchange of two antikinks comes from the ``Berry phase'' associated with the adiabatic motion of singlet bonds.  While this picture of fermionic spinons resembles some of the previous proposals inspired by fermionic large-$N$ expansions (e.g., Hastings\cite{PhysRevB.63.014413}), there are important differences.  We find that fermions experience a strong exchange-mediated attraction in the singlet channel, which causes them to form tightly bound pairs, thus invalidating the picture of a fermion sea with a Fermi surface or Dirac points.  We  also find that the emergent compact U(1) gauge field manifests itself not as a background magnetic flux, but as a quantized electric field of unit strength whose presence strongly constrains the motion of antikinks carrying a U(1) charge of $+2$ and even more strongly affects antikink pairs (charge $+4$).  It remains to be seen whether further progress can be made in the problem of spinons interacting with one another and the gauge field, a many-body problem with strong interactions.

In the present work, we calculate the dynamical structure factor of low-energy spin excitations in a kagome antiferromagnet observable by inelastic neutron scattering.  The physical process responsible for the lowest-energy magnetic scattering is the breaking up of a $S=0$ antikink pair into two antikinks with parallel spins that subsequently move away from each other.  If the ground state of the system is a valence-bond crystal the moving spinons disturb the preferred valence-bond arrangement.  The resulting energy increase leads to spinon confinement.  We neglect this effect because numerical diagonalization\cite{EuroPhysJB.2.501} and series expansion\cite{singh:180407} indicate that the energy differences between various valence-bond configurations are very small, on the order of $10^{-3} J$ per site.  We therefore expect that the confinement length is long and that its effects can be neglected in the first approximation.

\begin{figure}
    \includegraphics[width=0.95\columnwidth]{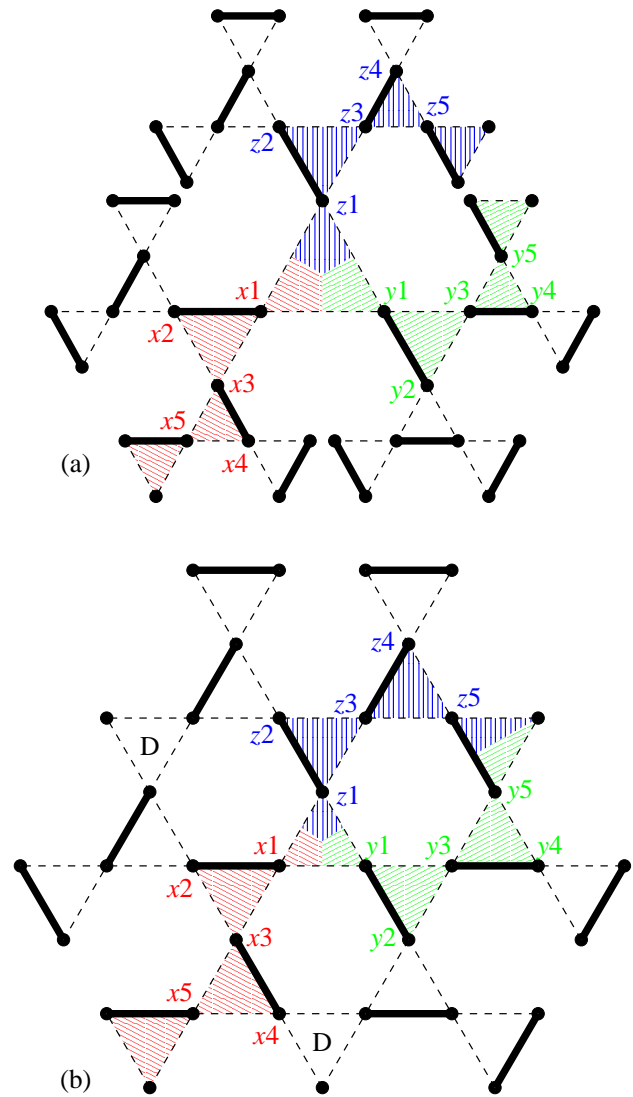}
  \caption{A correspondence between spinon trails on the Husimi cactus (a)  and kagome (b). Spins along the three shaded trails are labeled $\mathbf{S}_{\alpha n}$.  Here $\alpha=x,y,z$ denotes a trail and $n=1,2,3\ldots$ enumerates the spins along it.  The correspondence breaks down for $n > 5$ because trails on kagome may begin to overlap.  Further complications are brought by the presence of other defect triangles (labeled D) next to the trails.}\label{comp}
\end{figure}

We also neglect the influence of antikink pairs on one another.  Although the liberated antikinks may run into other pairs present on the lattice, this is not a severe problem.  The structure factor is determined by the overlap of the initial and final wavefunctions of the antikink pair.  Because the initial state is well localized, the result of the calculation is not sensitive to the long-distance behavior of the final state.  We compute the dynamical spin correlations in real space in the presence of a single antikink pair on the Husimi cactus.  The resulting correlation function decays quickly as we move away from the defect triangle.  Assuming that the spin correlations have similarly local nature on kagome proper, we translate the obtained spin correlations to kagome using the correspondence depicted in Fig. \ref{comp}.  Finally, after a spatial Fourier transform, we obtain the structure factor in 
$\mathbf k$ space at the edge of the spin gap.

The paper is organized as follows. In section \ref{COW}, we review the calculations of the wave functions of two antikinks with total spin $S=0$ and 1 on the Husimi cactus.  Section \ref{DSFC} describes the calculation of the dynamical structure factor. We discuss the results in section \ref{DAC}.

\section{Calculations of wave functions}\label{COW}
\begin{figure}
  \includegraphics[width=0.95\columnwidth]{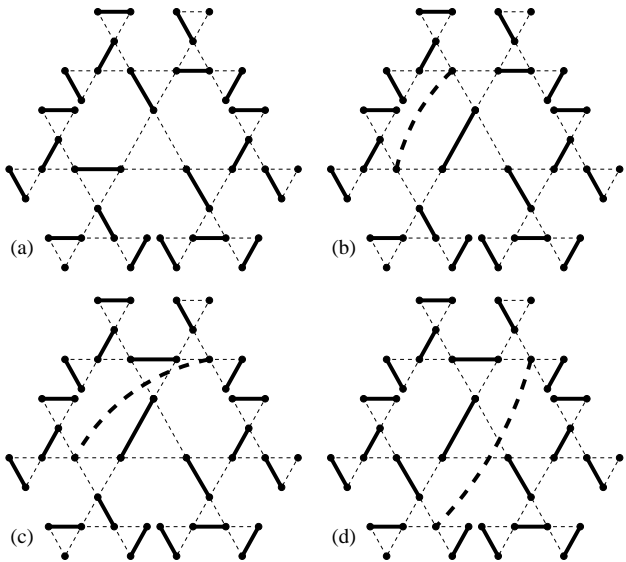}
  \caption{Valence-bond states of the Husimi cactus. The state with a single defect triangle (a) evolves into states with two antikinks (b-d) traveling through the cactus.  The dashed line indicates that the two antikinks have total spin 0.}\label{cactus}
\end{figure}

In order to study a defect triangle in isolation, Elser and Zeng\cite{PhysRevB.48.13647} examined kagome lattices in two-dimensional spaces of constant curvature where hexagonal loops of the familiar kagome are replaced with loops of length $L$.  Lattices with $L<6$ are finite and can be embedded in a two-dimensional sphere,\cite{rousochatzakis:094420} whereas lattices with $L>6$ are infinite and live in two-dimensional hyperbolic spaces.\cite{PhysRevB.48.13647}  The $L = \infty$ structure is a tree of corner-sharing triangles known as the Husimi cactus\cite{JStatPhys1.27.1237, JPhysA.5.1541} (Fig. \ref{comp}).  It has just the right ratio between the numbers of sites and links to permit the construction of a static valence-bond ground state and to study a defect triangle in isolation.  

A single defect triangle turns out to be a bound state of two antikink spinons with total spin 0.  The bound state is formed because spinons with antiparallel spins experience exchange-mediated attraction.  In contrast, two antikinks with total spin 1 repel and thus do not form a bound state.

\subsection{Two antikinks with $S=0$}

\subsubsection{Hilbert space}
\begin{figure}
  \includegraphics[width=0.95\columnwidth]{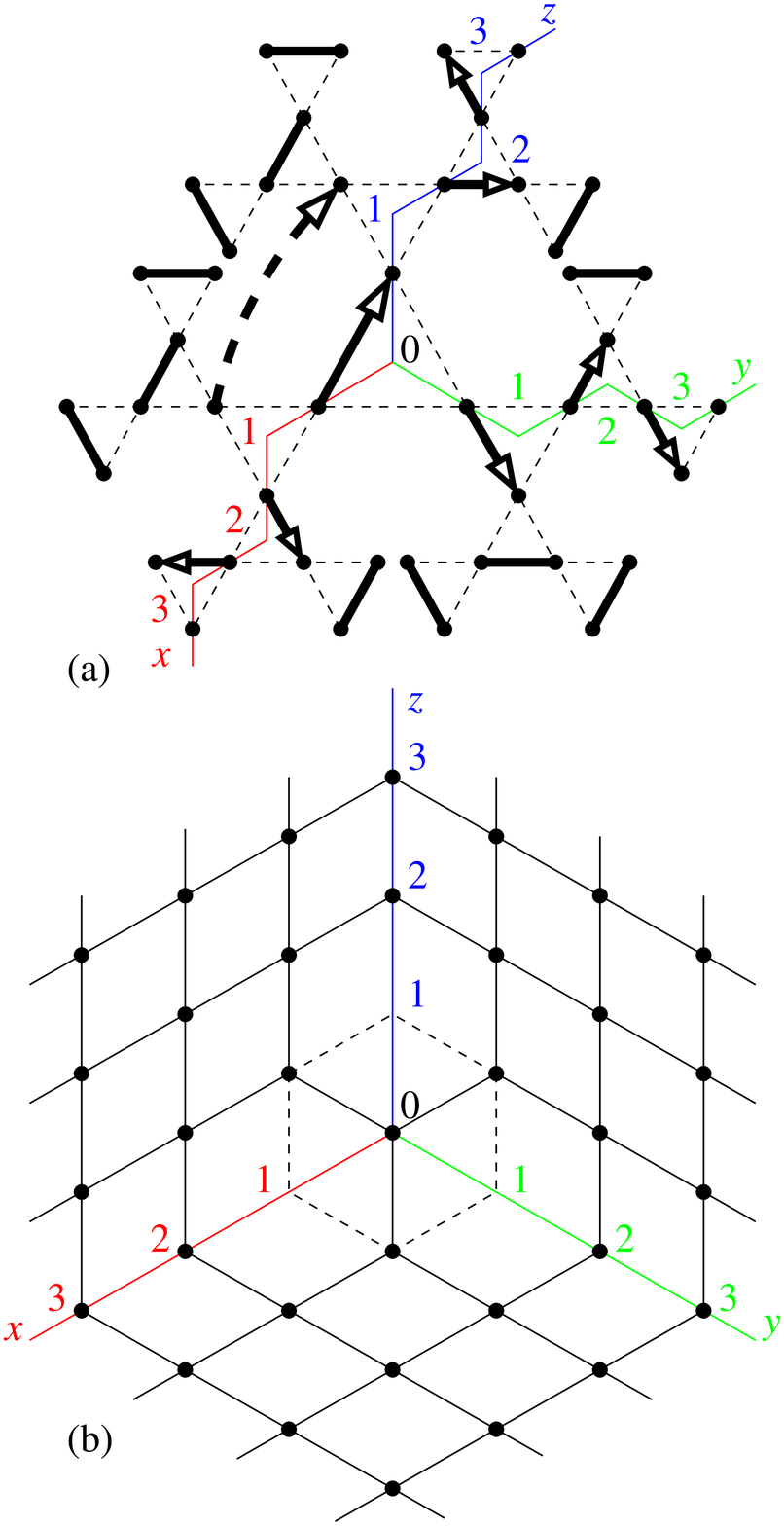}
  \caption{Hilbert space $A_2^{0}$ of two antikinks with $S=0$. (a) Two antikinks can move among three trails of the cactus labeled as $x$, $y$, and $z$ trails.  (b) The lattice of $|x,y,z\rangle$ states with $xyz = 0$ resembles the corner of a cube.}\label{corner}
\end{figure}

We begin by characterizing the Hilbert space to which the defect triangle state belongs. Upon applying the exchange Hamiltonian on the defect triangle state once, the defect triangle breaks into two defects of a new type that connected by a long range singlet: for example, if we apply left bond of the center trianlge in part a of Fig, the resulting state is shown in part b of Fig. These defects are close analogs of antikinks found in the one-dimensional sawtooth, or $\Delta$ chain.\cite{PhysRevB.53.6393,PhysRevB.53.6401}  They are domain walls separating the two distinct ground states of the $\Delta$ chain. The antikinks carry spin $1/2$ and are thus spinons.  Under the action of the exchange Hamiltonian, the two antikinks can move along three one-dimensional paths that meet at location of the defect triangle.  We will identify the three trails as $x$ $y$ and $z$ (Fig. \ref{corner}). The two antikinks can never be on the same trail.  A generic state can therefore be written as $|x,y,z\rangle$, where $x$, $y$, and $z$ are integers whose product vanishes.  For example, in the state $|2,3,0\rangle$, one of the antikinks is on the second triangle of the $x$ trail, while the other is on the third triangle of the $y$ trail.  Note that states $|0,0,1\rangle$ $|0,1,0\rangle$ or $|1,0,0\rangle$ are identical to $|0,0,0\rangle$ (the original defect triangle), so they can be omitted altogether.  It is the mathematical equivalence of the fact that any of the three singlets adjacent to the defect triangle can be viewed as a bound state of two anti-kinks.

The resulting Hilbert space is denoted $A_{2}^{0}$.  Generally, $A_{n}^S$ is the Hilbert space of $n$ antikinks with total spin $S$.  The Hilbert space $A_{2}^{0}$ is illustrated in Fig. \ref{corner}.

Two basis states $|\mathbf r\rangle \equiv |x,y,z\rangle$ and $|\mathbf r'\rangle \equiv |x',y',z'\rangle$ are not orthogonal to one another because they are not eigenstates of the same Hermitian operator.  Their overlap can be computed by following the standard recipe\cite{PhysRevB.37.3786}
\begin{equation}\label{overlap}
    \langle\mathbf r|\mathbf r'\rangle=\prod_{c}\epsilon_c \, 2^{1-L_c/2}
\end{equation}
where the product is over loops formed by superimposing if the dimer coverings of states $|\mathbf r\rangle$ and $|\mathbf r'\rangle$,  $L_c$ is the length of loop $c$, and $\epsilon_c = \pm 1$ is a $Z_2$ phase factor dependent on the sign convention for spin singlets.  The $S=0$ state of two spins on sites $i$ and $j$ is antisymmetric under exchange: $|(i,j)\rangle \equiv (|\uparrow_i \downarrow_j\rangle - |\downarrow_i \uparrow_j \rangle)/\sqrt{2} = -|(j,i)\rangle$.  To remove the ambiguity, the state $|(i,j)\rangle$ is shown as an arrow
pointing from $i$ to $j$.  The singlet phases need to be specified on the central triangle and along the three trails; all the other singlets are localized and their $Z_2$ phases are irrelevant. Because the three trails can be regarded as sawtooth chains, we choose the following phase convention.\cite{PhysRevLett.103.187203}  A singlet points from the base of a triangle on the sawtooth chain to its vertex. In states $|x,y,0\rangle$, $|0,y,z\rangle$ and $|x,0,z\rangle$, the long-range singlet has the same direction with the singlet on the center triangle, as shown in Fig. \ref{corner}. In states $|x,0,0\rangle$, $|0,y,0\rangle$ and $|0,0,z\rangle$, the long-range singlet points from the central triangle to the triangle on one of the trails.  Overlaps of various valence-bond states have the following values:
\begin{subequations}
\begin{eqnarray}
  \langle x,y,0|x',y',0\rangle &=& 2^{-|x-x'|-|y-y'|},\\
  \langle x,0,0|0,y,z\rangle &=& 2^{2-x-y-z},\\
  \langle x,y,0|x',0,z'\rangle &=& 2^{-|x-x'|-y-z'},\\ 
  \langle x,y,0|0,0,0\rangle &=& 2^{1-x-y},\\
  \langle x,0,0|0,0,0\rangle &=& 2^{1-x}.
\end{eqnarray}
The rest of the overlaps can be obtained by using permutations of $x$, $y$, and $z$. 
\label{eq:overlaps}
\end{subequations}

With two antikinks in the three trails, at least one of the ``coordinates'' $x$, $y$, $z$ is zero.  For that reason, the motion of two antikinks can be mapped onto the motion of a single particle on three faces of an infinite cube, Fig. \ref{corner} (b).  This mapping reveals a simple rule that can be used to determine the overlap in all cases described in Eq.~(\ref{eq:overlaps}).  Let $n(\mathbf r, \mathbf r')$ be the length of the shortest path (expressed as the number of links) that connects points $\mathbf r$ and $\mathbf r'$ in Fig. \ref{corner} (b).  The overlap between the two states is
\begin{equation}
    \langle \mathbf r|\mathbf r'\rangle=2^{-n(\mathbf r, \mathbf r')}.
\end{equation}

\subsubsection{Orthonormal basis}
The basis states $|x,y,z\rangle$ can be orthogonalized by performing a simple rotation. We will denote the orthogonalized basis states as $|x,y,z\rangle^\mathrm{o}$ and the orthonormalized states as $|x,y,z\rangle^{\mathrm{on}}$. In the orthogonalization procedure, the defect triangle state remains the same:
\begin{subequations}
\begin{equation}
|0,0,0\rangle^\mathrm{o} = |0,0,0\rangle. \label{eq:orth-000}
\end{equation}
Its nearest neighbors are transformed as follows:
\begin{eqnarray}
|1,1,0\rangle^\mathrm{o}
= |1,1,0\rangle - \frac{1}{2}|0,0,0\rangle. \label{eq:orth-011}
\end{eqnarray}
The resulting states $|0,1,1\rangle^\mathrm{o}$,
$|1,0,1\rangle^\mathrm{o}$, and $|1,1,0\rangle^\mathrm{o}$ are
orthogonal to $|0,0,0\rangle^\mathrm{o}$ and to one other.

The transformation for states along cube edges is:
\begin{equation}
|x,0,0\rangle^\mathrm{o}
= |x,0,0\rangle - \frac{1}{2}|x-1,0,0\rangle. \label{eq:orth-x00}
\end{equation}
The resulting state $|x,0,0\rangle^\mathrm{o}$ is orthogonal to any
state $|0,y,z\rangle$ with no antikink in the $x$ trail.  It is
also orthogonal to any state $|x',y,0\rangle$ with an antikink
closer to the origin, $x'<x$.  From that it follows that edge states
$|x,0,0\rangle^\mathrm{o}$, $|0,y,0\rangle^\mathrm{o}$, and
$|0,0,z\rangle^\mathrm{o}$ are orthogonal to one another and to the
four states with antikinks near the origin.

Finally, states $|x,y,0\rangle$ with antikinks away from the origin
($x, y > 1$) are transformed by combining translations in two
directions:
\begin{eqnarray}
|x,y,0\rangle^\mathrm{o} &=& |x,y,0\rangle -
\frac{1}{2}|x-1,y,0\rangle \label{eq:orth-xy0}
\\
&&- \frac{1}{2}|x,y-1,0\rangle + \frac{1}{4}|x-1,y-1,0\rangle.
\nonumber
\end{eqnarray}
\label{eq:singlet-o}
\end{subequations}

The orthonormal basis $\{|x,y,0\rangle^\mathrm{on}\}$ is now easily obtained by normalization:
\begin{subequations}
\begin{eqnarray}
|0,0,0\rangle^\mathrm{on} &=& |0,0,0\rangle^\mathrm{o},\\
|1,1,0\rangle^\mathrm{on} &=& \frac{2}{\sqrt{3}}|1,1,0\rangle^\mathrm{o},\\
|x,0,0\rangle^\mathrm{on} &=& \frac{2}{\sqrt{3}}|x,0,0\rangle^\mathrm{o},\\
|x,y,0\rangle^\mathrm{on} &=& \frac{4}{3}|x,y,0\rangle^\mathrm{o}.
\end{eqnarray}
\label{eq:singlet-on}
\end{subequations}

\subsubsection{Effective Hamiltonian and the spectrum}
 
In the bulk, the Hamiltonian matrix is very simple: 
\begin{widetext}
\begin{subequations}
\begin{equation}\label{hamimatrix}
    H|x,y,0\rangle^\mathrm{on}=
    -\frac{J}{2}|x+1,y,0\rangle^\mathrm{on}
    -\frac{J}{2}|x-1,y,0\rangle^\mathrm{on}
    -\frac{J}{2}|x,y+1,0\rangle^\mathrm{on}
    -\frac{J}{2}|x,y-1,0\rangle^\mathrm{on}
    +\frac{5J}{2}|x,y,0\rangle^\mathrm{on}
\end{equation}
The Hamiltonian can be loosely understood as follows: two particles(antikinks) hop on the three one dimensional trails with $-J/2$ as their hopping amplitude and an attractive short-range potential.  (We measure the energy from its value in the ground state.)

The action of the Hamiltonian on the state with a defect triangle is as follows:
\begin{equation}\label{defect}
    H|0,0,0\rangle^\mathrm{on}=
    -\frac{\sqrt{3}J}{4}|1,1,0\rangle^\mathrm{on}
    -\frac{\sqrt{3}J}{4}|1,0,1\rangle^\mathrm{on}
    -\frac{\sqrt{3}J}{4}|0,1,1\rangle^\mathrm{on}
    +\frac{3J}{4}|0,0,0\rangle^\mathrm{on}.
\end{equation}
(Note the reduced diagonal term.)

For a state with one of the antikinks next to the defect triangle and the other farther away, $|x,1,0\rangle^\mathrm{on}$ with $x>1$, $y=1$, we have
\begin{equation}\label{adjacent}
    H|x,1,0\rangle^\mathrm{on}=
    -\frac{J}{2}|x+1,1,0\rangle^\mathrm{on}
    -\frac{J}{2}|x-1,1,0\rangle^\mathrm{on}
    -\frac{J}{2}|x,2,0\rangle^\mathrm{on}
    -\frac{\sqrt{3}J}{4}|x,0,0\rangle^\mathrm{on}
    +\frac{5J}{2}|x,1,0\rangle^\mathrm{on}
    +\frac{J}{4}|x,0,1\rangle^\mathrm{on}.
\end{equation}
For a state with one antikink on the original defect triangle and the other somewhere in trail $x$, $|x,0,0\rangle^{\mathrm{on}}$,
 \begin{equation}\label{axis}
    H|x,0,0\rangle^\mathrm{on} =a
    -\frac{J}{2}|x+1,0,0\rangle^\mathrm{on}
    -\frac{J}{2}|x-1,0,0\rangle^\mathrm{on}
    -\frac{\sqrt{3}J}{4}|x,1,0\rangle^\mathrm{on}
    -\frac{\sqrt{3}J}{4}|x,0,1\rangle^\mathrm{on}
    +2J|x,0,0\rangle^\mathrm{on}.
 \end{equation}
\label{eq:H-singlet}
\end{subequations}
 \end{widetext}
Other special cases are considered in the supplemental material of Ref. \onlinecite{PhysRevLett.103.187203}.

To determine the energy spectrum, we truncated the cactus at a finite radius $R$ and diagonalized the Hamiltonian numerically. The spectrum for a singlet pair of antikinks consists of a two-particle continuum starting at $E=J/2$ and a non-degenerate bound state at $E=0.44J$, i.e. $0.06J$ below the bottom of the continuum. The diameter of the bound state, obtained by fitting its energy $E(R)$ to the form $E+\delta E\exp(-2R/\xi)$, is $\xi = 2.8$  lattice spacings.  The probability to find both antikinks on the same triangle is 0.72, which implies a tight bound state.

\subsection{Two antikinks with $S=1$}
The energy spectrum of two antikinks in a $S=1$ state was computed in the same fashion.  We mapped out the Hilbert space $A_2^{1}$ for a triplet pair by breaking one of the three singlets adjacent to the defect triangle at the center and letting the resulting spinons propagate along the three trails. The basis states, also be denoted $|x,y,z\rangle$, are again non-orthogonal. This time, there is no $|0,0,0\rangle$ state, while $|1,0,0\rangle$, $|0,1,0\rangle$ and $|0,0,1\rangle$ states do exist and are physically distinct. Here are some examples of the overlap matrix elements between the states in $A_2^1$:
\begin{subequations}
\begin{eqnarray}
  \langle x_1,y_1,0|0,y_2,0\rangle &=& \left(\frac{1}{2}\right)^{x_1+|y_1-y_2|}, \\
   \langle x_1,y_1,0|x_2,0,0\rangle&=& -\left(\frac{1}{2}\right)^{y_1+|x_1-x_2|}, \\
  \langle x_1,y_1,0|x_2,y_2,0\rangle &=& \left(\frac{1}{2}\right)^{|x_1-x_2|+|y_1-y_2|}, \\
  \langle x_1,y_1,0|0,0,z_2\rangle &=& 0, \\
  \langle x_1,y_1,0|0,y_2,z_2\rangle &=& - \left(\frac{1}{2}\right)^{x_1+|y_1-y_2|+z_2}.
\end{eqnarray}
\label{eq:triplet-overlap}
\end{subequations}
Other cases can be obtained by making use of the permutational symmetry.

\begin{figure}
  \includegraphics[width=0.95\columnwidth]{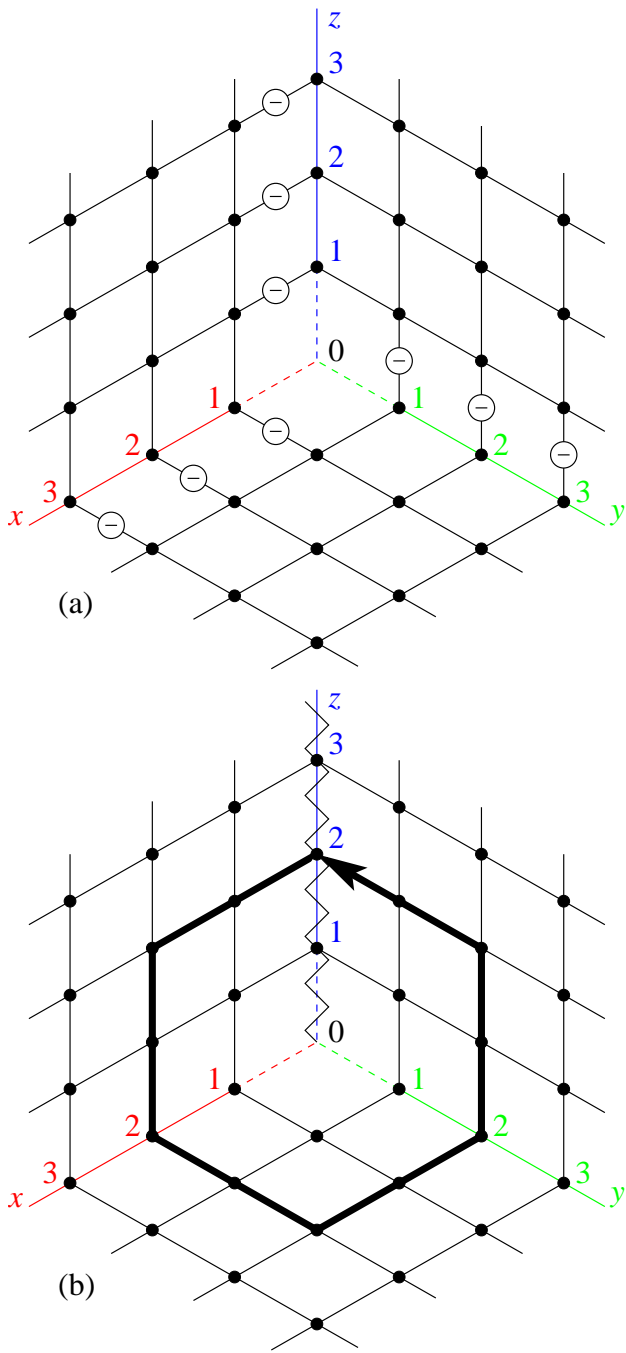}
  \caption{Hilbert space $A_2^1$ of two antikinks with $S=1$.  (a) Minus signs denote bonds with the inverted sign of the hopping amplitude, $t>0$. (b) Making a loop around the center changes sign of the wave function.}\label{a21}
\end{figure}

A similar procedure can be followed to orthogonalize most of the states. However, we did not find an obvious way to orthogonalize the states with spinons next to the original defect triangle, namely $|1,0,0\rangle$, $|0,1,0\rangle$, $|0,0,1\rangle$, $|1,1,0\rangle$, $|1,0,1\rangle$ and $|0,1,1\rangle$. We orthogonalize the rest of the basis states as exemplified by
\begin{equation}\label{orthogonalize}
\begin{aligned}
    |x,1,0\rangle^{\mathrm{o}}=&|x,1,0\rangle+\frac{1}{2}|x,0,0\rangle\\
&-\frac{1}{2}|x-1,1,0\rangle-\frac{1}{4}|x-1,0,0\rangle.
    \end{aligned}
\end{equation}
Because some of the basis states are not orthogonal, the energy spectrum is obtained by solving the generalized eigenvalue problem: $H \psi = E O \psi$ where $H$ is the Hamiltonian matrix, $O$ is the overlap matrix and $\psi$ is an eigenvector.

The low-energy effective Hamiltonian is similar to that of the $S=0$ sector, with two important differences. First, the hopping amplitude between two states with a negative overlap is positive instead of negative. Most of these ``wrong'' signs can be corrected by a gauge transformation. However, the total phase accumulated on a loop containing the center in Fig.~\ref{a21} adds up to a flux $\pi$, which reflects the fermionic nature of the antikinks.  Second, in the singlet case the ``potential'' energy of the two antikinks is lowered by $3J/4$ when they both reside on the central triangle. In contrast, antikinks with parallel spins repel each other: the energy of the system is raised by $J/4$ when they share the central triangle.  Consequently, the spectrum of a triplet pair of antikinks consists of a two particle continuum starting at $E = J/2$ with no bound states. The ground state in the $S=1$ sector is doubly degenerate.   Both of these ground states have a line of nodes.  The symmetry and degeneracy of the ground states are consistent with the Fermi statistics of antikinks.

\section{Dynamical structure factor calculation}\label{DSFC}
The real-space dynamical structure factor is defined as
\begin{equation}\label{dsfreal}
    S(\omega,\mathbf{R},\mathbf{R}^{\prime})=\sum_{f}\delta(E_f-E_i - \hbar\omega)\langle i|S_{\mathbf{R}}^{-}|f\rangle\langle f|S_{\mathbf{R}'}^{+}|i\rangle.
\end{equation}
Inelastic neutron scattering directly measures its Fourier transform, 
\begin{equation}\label{Sq-FT}
S_N(\omega,\mathbf{q}) = \sum_{f}\delta(E_f-E_i-\hbar\omega)
\left|\sum_{\mathbf R} \langle f|S_{\mathbf R}^{+}|i\rangle
	e^{i\mathbf{q}\cdot\mathbf{R}}\right|^2.
\end{equation}
In both the initial state $|i\rangle$ and in final states $|f\rangle$, the motion of the antikinks is restricted to three one-dimensional paths.  Singlet bonds along these trails shift as the spinons move past them.  In contrast, singlet bonds off the trails remain stationary.  If $\mathbf R$ is located off a spinon trail, states $S_{\mathbf{R}}^{+}|i\rangle$ and $|f\rangle$ are orthogonal because the former has a triplet on a bond involving site $\mathbf R$, whereas the latter has a singlet there.  For that reason, the structure factor (\ref{dsfreal}) is nonzero only when both $\mathbf R$ and $\mathbf R'$ are on the spinon trails.  Spins along the trail will be labeled $\mathbf{S}_{\alpha n}$, where $\alpha = x$, $y$, or $z$ is the trail index and $n=1,2,3\ldots$ is the position along the trail (Fig. \ref{comp}).  $\mathbf{R}_{\alpha n}$ is the physical location of that spin on kagome.

We evaluate the real-space structure factor (\ref{dsfreal}) on the cactus for the lowest energy transfer equal to the binding energy of an antikink pair.  These dynamic correlations turn out to be sufficiently short-ranged to justify the extrapolation of the result to a pair of antikinks on kagome via the correspondence illustrated in Fig.~\ref{comp}.  We assume that antikink pairs are randomly distributed over the lattice and therefore the net scattering intensity is the sum of scattering intensities of individual pairs (no interference).   Because different pairs live in different valence-bond backgrounds, we average the structure factor over all possible configurations of the spinon trails.  
\begin{figure}
  \flushleft
  \includegraphics[width=0.49\columnwidth]{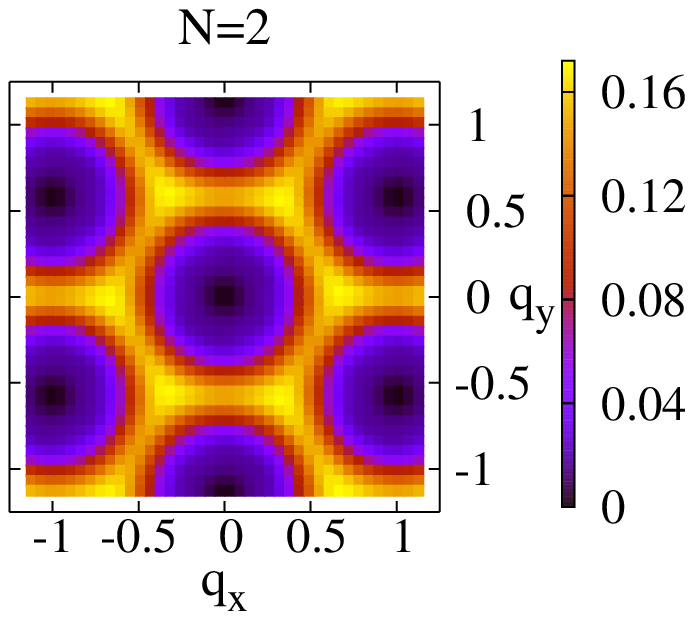}
  \includegraphics[width=0.49\columnwidth]{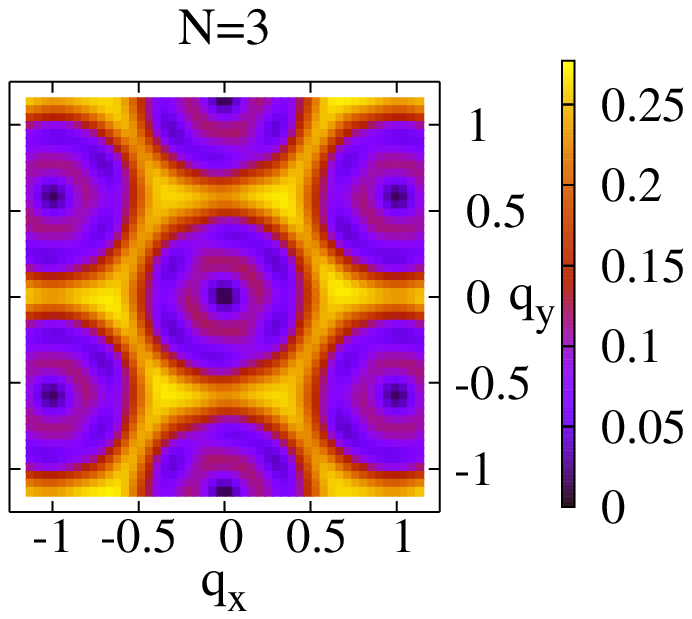}
  
  \medskip
  \includegraphics[width=0.49\columnwidth]{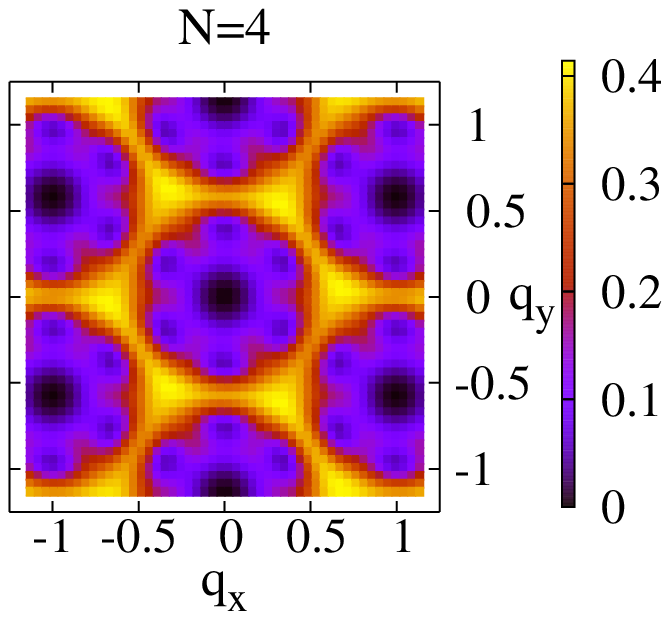}
  \includegraphics[width=0.49\columnwidth]{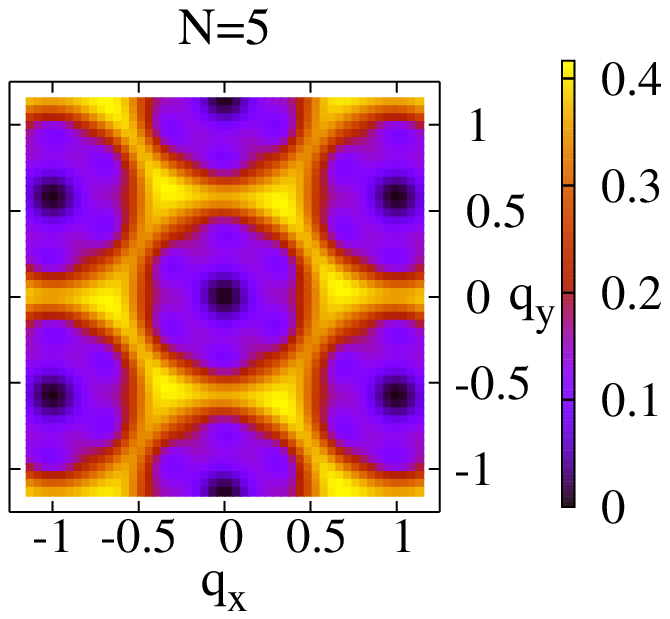}
  
  \medskip
  \includegraphics[width=0.49\columnwidth]{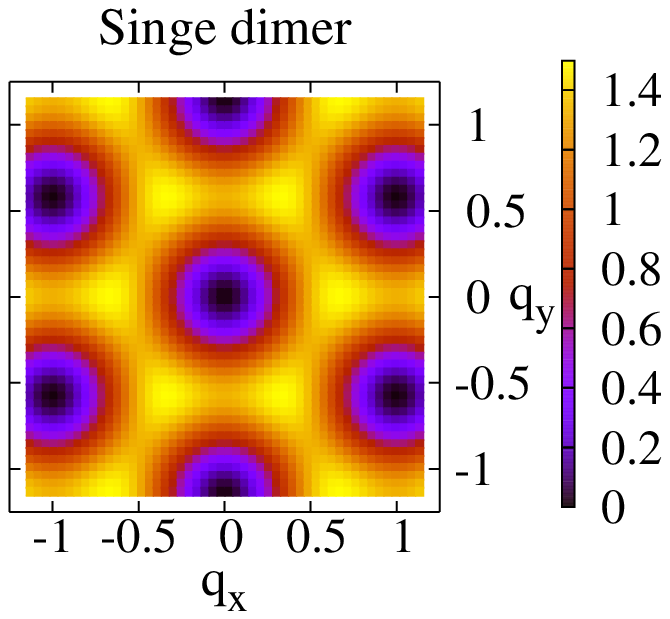}  
  \caption{The computed dynamical structure factor $N=2$, 3, 4, and 5 as well
  as for a few isolated dimers randomly distributed on the lattice.  Wavenumbers are measured in units of $2\pi/a$, where $a$ is the lattice constant.
 }
\label{dyna}
\end{figure}

To simplify the calculations, we only approximate the singlet ground state as a superposition of four states  $|0,0,0;\mathrm{s}\rangle$ (the defect triangle), $|1,1,0;\mathrm{s}\rangle$, $|0,1,1;\mathrm{s}\rangle$ and $|1,0,1;\mathrm{s}\rangle$ (spinons hop to adjacent triangles).  The truncated ground state has the overlap of $0.92$ with the actual one.  No truncation is done for the final triplet states, in which the spinons are delocalized.  

The form factor now reduces to
\begin{eqnarray}\label{Sq}
S_N(\omega,\mathbf{q}) &=& \sum_{f}\delta(E_f-E_i-\hbar\omega)
\\
&\times& \left|\sum_{\alpha} \sum_{n=1}^N \langle f|S_{\alpha n}^{+}|i\rangle
	\exp{(i\mathbf{q}\cdot\mathbf{R}_{\alpha n})}\right|^2,
\nonumber
\end{eqnarray}
In this expression, we include spins $\mathbf S_{\alpha n}$ located along the trails with $n \leq N$. In theory, one would like to take the limit $N \to \infty$. In practice, we can only go to $N=5$; for larger values of $N$ trails may start overlapping on kagome.  Fortunately, it can be seen in Fig.~\ref{dyna} that the structure factor changes little between $N=4$ and $N=5$, so we use the $N=5$ result as our final answer.

The matrix elements $\langle f|S_{\alpha n}^{+}|i \rangle$ are calculated as follows.  The initial singlet ground state is a linear combination of four states $|x,y,z;\mathrm{s}\rangle$ from the singlet space $A_{2}^{0}$. The final states are linear combinations of $|x,y,z;\mathrm{t}\rangle$ in the triplet space $A_{2}^{1}$.  (As in previous sections, we use the shorthand $\mathbf r$ for the three ``coordinates'' of the antikink pair $x$,$y$ and $z$.  The physical coordinates of a site on kagome are denoted $\mathbf R_{\alpha n}$.)  We first expand the initial and final states in the non-orthogonal singlet basis $\{|\mathbf{r};\mathrm{s}\rangle\}$ and partially orthogonalized triplet basis $\{|\mathbf{r};\mathrm{t}\rangle^\mathrm{on}\}$:
\begin{subequations}
\begin{eqnarray}
    |i\rangle &=& \sum_{\mathbf{r}}c_{\mathbf{r}}(\mathrm{s})|\mathbf{r};\mathrm{s}\rangle,\\
    |f\rangle &=& \sum_{\mathbf{r}}c_{\mathbf{r},f}(\mathrm{t})|\mathbf{r};\mathrm{t}\rangle^{\mathrm{on}},
\end{eqnarray}
\label{eq:states-i-and-f}
\end{subequations}
where $f$ labels the two final states.  After a computation of matrix elements
\begin{equation}\label{matrixE}
    M_{\alpha n}(\mathbf{r},\mathbf{r}')\equiv {}^{\mathrm{on}}\langle \mathbf{r};\mathrm{t}|S^{+}_{\alpha n}|\mathbf{r}';\mathrm{s}\rangle,
\end{equation}
listed in the Appendix, we obtain
\begin{equation}
\langle f|S_{\alpha n}^{+}|i\rangle = \sum_{\mathbf r, \mathbf r'}
c^*_{\mathbf{r},f}(\mathrm{t}) M_{\alpha n}(\mathbf{r},\mathbf{r}')
	c_{\mathbf{r}}(\mathrm{s}),
\end{equation}
and use it in Eq.~(\ref{Sq}).

The dynamical structure factor is displayed in Fig. \ref{dyna}.

\section{Discussion}\label{DAC}

We have computed the $\mathbf q$ dependence of the dynamical structure factor $S(\mathbf q, \omega)$ at the edge of the spin gap, $\Delta = 0.06J$.  Excitations responsible for this spectral weight are pairs of antikinks with parallel spins freed from their $S=0$ bound state.  The wavevector dependence of the dynamical structure factor resembles that of  isolated pairs of spins interacting via Heisenberg exchange:
\begin{equation}\label{}
S(\omega,\mathbf{q}) = \sum_{\mathbf a}
   	2[1-\cos(\mathbf{q}\cdot\mathbf{a})]\delta(\hbar\omega - J),
\end{equation}
where $\mathbf{a}$ are separations of spins within a pair.  For three dimer orientations, as on kagome, $S(\omega,\mathbf{q})$ is shown in Fig.~\ref{dyna} next to our result for the structure factor resulting from the breaking of antikink pairs.  The resemblance is not surprising because the ground state of the Heisenberg model on kagome can be pictured as a collection of slowly resonating valence bonds.  The structure factor of the inelastic neutron scattering in powder herbertsmithite indeed bears resemblance to that of isolated dimers.\cite{PhysRevLett.103.237201}  A similar $\mathbf q$ dependence was found for the instantaneous (frequency-integrated) structure factor by Laeuchli and Lhuillier\cite{laeuchli-2009} by using exact diagonalization.  More recently, Singh\cite{singh-neutrons} computed the dynamical structure factor due to another kind of excitations, viz. creation of kink-antikink pairs with a larger onset energy, $\Delta' = 0.25J$, and found a similar intensity distribution.  

Our calculation is based on several assumptions.  (1) We neglect the effects of virtual excitations in the form of kink-antikink pairs near a moving antikink.  Our previous study\cite{PhysRevLett.103.187203} of a related one-dimensional system (the $\Delta$ chain\cite{PhysRevB.53.6393, PhysRevB.53.6401}) showed that these excitations create a small renormalization of the antikink parameters but are otherwise harmless.  (2) We assumed that the bound  pairs of two antikinks with $S=0$ are sufficiently small so that their properties on a periodic kagome lattice and its tree-like analog are essentially the same.  The small diameter of a pair, $\xi = 2.8$ lattice spacings, provides assurance that this approximation is not unreasonable.  (3) We neglected the dynamics of the pairs in the ground state of the Heisenberg model on kagome.  This assumption is justified if the pairs move slowly on the time scale of the inverse spin gap.  Some indications that this is so come from numerical work indicating a large density of singlet states at low energy.\cite{EuroPhysJB.2.501}  Series expansion also indicates that the energy splittings between low-lying singlet states are very small, with energy differences as small as $10^{-3} J$ per site.\cite{singh:144415}  (4) We assumed that pair positions exhibit no long-range order and thus their scattering amplitudes add incoherently.  This may not be the case if the system indeed has valence-bond order, which requires a periodic arrangement of spinon pairs in the ground state.  However, the closeness of energy levels in the singlet sector suggests that the valence-bond crystal is fragile and can easily turn into a disordered solid under the influence of bond disorder or nonmagnetic impurities.\cite{PhysRevB.68.224416, singh-neutrons}  (5) We ignored the possibility of interactions between adjacent pairs of antikinks.  The dynamics of antikinks may be altered if their trails pass close to another defect triangle (Fig.~\ref{comp}).  This is a many-body problem that we hope to address in future work.

\section*{Acknowledgments}

We thank Collin Broholm for stimulating this work and Fr\'ed\'eric Mila and Rajiv Singh for helpful discussions.  This project was supported by the U.S. Department of Energy, Office of Basic Energy Sciences, Division of Materials Sciences and Engineering under Award DE-FG02-08ER46544.

\appendix
\section{Matrix elements}\label{ME}
Before we give a list of matrix elements $M_{\alpha n}(\mathbf{r};\mathbf{r}')$, we need to specify a convention of labeling all the spins. The three one-dimensional trails on which the antikinks are moving can be viewed as three sawtooth chains originating at the center triangle. We name a spin using the trail it's on and the distance between it and the center triangle. The three vertices of the center triangle are labeled $\mathbf{S}_{x1}$, $\mathbf{S}_{y1}$ and $\mathbf{S}_{z1}$, Fig. \ref{comp}.

Since we write the initial state a linear combination of $|0,0,0;\mathrm{s}\rangle$, $|1,1,0;\mathrm{s}\rangle$, $|0,1,1;\mathrm{s}\rangle$ and $|1,0,1;\mathrm{s}\rangle$ and our convention for basis states respects the symmetry of permutations among ``coordinates'' $x$, $y$, and $z$, we only list nonzero matrix elements involving states $|0,0,0;\mathrm{s}\rangle$ and $|1,1,0;\mathrm{s}\rangle$.  For $|0,0,0;\mathrm{s}\rangle$,
\begin{eqnarray}
  \langle 1,0,0;\mathrm{t}|S_{x1}^{+}|0,0,0;\mathrm{s}\rangle &=& -\frac{1}{\sqrt{2}}, \nonumber\\
  \langle 0,1,0;\mathrm{t}|S_{y1}^{+}|0,0,0;\mathrm{s}\rangle &=& -\frac{1}{\sqrt{2}}, \nonumber\\
  \langle 0,0,1;\mathrm{t}|S_{z1}^{+}|0,0,0;\mathrm{s}\rangle &=& -\frac{1}{\sqrt{2}}, \nonumber\\
  {}^{\mathrm{on}}\langle n,0,0;\mathrm{t}|S_{x,2n-1}^{+}|0,0,0;\mathrm{s}\rangle &=& -\frac{1}{\sqrt{6}}\left(\frac{1}{2}\right)^{n-2}, \nonumber \\
  {}^{\mathrm{on}}\langle 0,n,0;\mathrm{t}|S_{y,2n-1}^{+}|0,0,0;\mathrm{s}\rangle &=& -\frac{1}{\sqrt{6}}\left(\frac{1}{2}\right)^{n-2}, \nonumber\\
  {}^{\mathrm{on}}\langle 0,0,n;\mathrm{t}|S_{z,2n-1}^{+}|0,0,0;\mathrm{s}\rangle &=& -\frac{1}{\sqrt{6}}\left(\frac{1}{2}\right)^{n-2},
\label{eq:App1}
\end{eqnarray}
with $n\ge 2$.  Similarly for $|1,1,0;\mathrm{s}\rangle$,
\begin{eqnarray}
  {}^{\mathrm{on}}\langle n,1,0;\mathrm{t}|S^+_{x,2n-1}|1,1,0;\mathrm{s}\rangle&=&\frac{1}{\sqrt{2}}\left(\frac{1}{2}\right)^{n}, \nonumber\\
  {}^{\mathrm{on}}\langle n,0,0;\mathrm{t}|S^+_{x,2n-1}|1,1,0;\mathrm{s}\rangle&=&-\frac{1}{\sqrt{6}}\left(\frac{1}{2}\right)^{n-1}, \nonumber\\
  {}^{\mathrm{on}}\langle 1,n,0;\mathrm{t}|S^+_{y,2n-1}|1,1,0;\mathrm{s}\rangle&=&-\frac{1}{\sqrt{2}}\left(\frac{1}{2}\right)^{n}, \nonumber\\
   {}^{\mathrm{on}}\langle 0,n,0;\mathrm{t}|S^+_{y,2n-1}|1,1,0;\mathrm{s}\rangle&=&-\frac{1}{\sqrt{6}}\left(\frac{1}{2}\right)^{n-1}, \nonumber  \\
   {}^{\mathrm{on}}\langle 0,0,n;\mathrm{t}|S^+_{z,2n-1}|1,1,0;\mathrm{s}\rangle&=&-\frac{1}{\sqrt{6}}\left(\frac{1}{2}\right)^{n-1},\nonumber\\  
   \langle 1,0,0;\mathrm{t}|S_{x1}^+|1,1,0;\mathrm{s}\rangle &=&-\frac{1}{2\sqrt{2}}, \nonumber\\
   \langle0,1,0;\mathrm{t}|S_{x1}^+|1,1,0;\mathrm{s}\rangle &=&\frac{1}{2\sqrt{2}}, \nonumber\\
   \langle1,0,1;\mathrm{t}|S_{x1}^+|1,1,0;\mathrm{s}\rangle &=&-\frac{1}{4\sqrt{2}}, \nonumber\\
   \langle0,1,1;\mathrm{t}|S_{x1}^+|1,1,0;\mathrm{s}\rangle &=&-\frac{1}{4\sqrt{2}}, \nonumber\\
   \langle1,0,0;\mathrm{t}|S_{x2}^+|1,1,0;\mathrm{s}\rangle &=&\frac{1}{2\sqrt{2}}, \nonumber\\
   \langle0,1,0;\mathrm{t}|S_{x2}^+|1,1,0;\mathrm{s}\rangle &=&-\frac{1}{2\sqrt{2}}, \nonumber\\
   \langle1,0,1;\mathrm{t}|S_{x2}^+|1,1,0;\mathrm{s}\rangle &=&\frac{1}{4\sqrt{2}}, \nonumber\\
   \langle0,1,1;\mathrm{t}|S_{x2}^+|1,1,0;\mathrm{s}\rangle &=&\frac{1}{2\sqrt{2}}, \nonumber\\
   \langle1,1,0;\mathrm{t}|S_{x2}^+|1,1,0;\mathrm{s}\rangle &=&-\frac{1}{\sqrt{2}}.
\label{eq:App3}
\end{eqnarray}

\bibliography{kagome}
\end{document}